\documentclass[useAMS,usenatbib]{mn2e}
\usepackage{lscape}
\usepackage{rotating}
\usepackage{amssymb}
\usepackage{float}
\usepackage{txfonts}
\usepackage{natbib}

\def\cm3{\hbox{cm$^{-3}$}}

\input psfig.sty
\input epsf.sty
\usepackage{graphicx}
\usepackage{epsfig}
\title[Testing the KYNREFREV model]
{Testing the X-ray reverberation model KYNREFREV in a sample of Seyfert 1 Active Galactic Nuclei}

\author[Caballero-Garc\'{i}a et~al.]{M.~D. Caballero-Garc\'{i}a$^1$\thanks{E-mail:
garcia@asu.cas.cz}, I.~E. Papadakis$^{2,3}$, M. Dov\v{c}iak$^{1}$, M. Bursa$^{1}$, \newauthor 
A. Epitropakis$^{2}$, V. Karas$^{1}$, J. Svoboda$^{1}$
\\
\\
$^{1}$ Astronomical Institute, Academy of Sciences, Bo\v{c}n\'{\i}~II~1401, CZ-14100~Prague, Czech~Republic \\ 
$^{2}$ Department of Physics and Institute of Theoretical and Computational Physics, University of Crete, 71003 Heraklion, Greece \\
$^{3}$ Foundation for Research and Technology - Hellas, IESL, Voutes, GR-7110 Heraklion, Greece \\
}

\date{Accepted. Received; in original form}
\pagerange{\pageref{firstpage}--\pageref{lastpage}}

\begin{document}
\maketitle
\label{firstpage}
\begin{abstract}
We present the first results obtained by the application of the KYNREFREV-reverberation model, which is ready for its use in XSPEC. This model 
computes the time dependent reflection spectra of the disc as a response to a flash of primary power-law radiation from a point source corona 
located on the axis of the black hole accretion disc (lamp-post geometry). Full relativistic effects are taken into account. The ionisation of 
the disc is set for each radius according to the amount of the incident primary flux and the density of the accretion disc. We tested the model 
by fitting model predictions to the observed time-lag spectra of three Narrow-Line Seyfert 1 galaxies (ARK~564, MCG-6-30-15 and 
1H~0707-495), assuming either a rapidly or zero spinning black hole (BH). The time-lags strongly suggest a compact X-ray source, located close 
to the BH, at a height of $\sim$ 4 gravitational radii. This result does not depend either on the BH spin or the disc ionization. There
is no significant statistical difference between the quality of the best-fits in the rapidly and zero spinning BH scenarios in Ark~564 and 
MCG-6-30-15. But there is an indication that the hypothesis of a non-rotating BH in 1H~0707-495 is not consistent with its time-lag 
spectrum. Finally, the best-fits to the Ark~564 and 1H~0707-495 data are of rather low quality. We detect wavy-residuals around the best-fit
reverberation model time-lags at high frequencies. This result suggests that the simple lamp-post geometry does
not fully explain the X-ray source/disc configuration in Active Galactic Nuclei.
\end{abstract}

\section{Introduction}
\label{sec:intro}

The widely accepted picture of the origin of the X-ray emission from Active Galactic Nuclei (AGN) is that soft thermal photons emitted from the accretion disc are upscattered to keV energies by hot electrons in a region which is commonly referred to as the X--ray ``corona''. The upscattered X-ray photons form a power-law shaped continuum spectrum (with a photon index of ${\Gamma}\sim2$), often called the primary X-ray radiation. Part of the power-law emission may be reflected from distant matter (torus or clouds in the Broad Line Region -- BLR)  or by the accretion disc, forming the additional ``reflection component'' in the overall spectrum. The main features of this component are:
(i) The Fe K$_{\alpha}$ line in the 6-7\,keV band (depending on the ionization of the reflecting material) which may be broadened, if reflection arises from the inner accretion disc. In this case the line will be relativistically smeared and its shape can be used as a diagnostic of the geometry of the accretion flow (e.g. \citealt{fabian99}). (ii) X--ray photons with energy larger than 50--100 keV are Compton downscattered, and produce an excess of photons in the 10--50~keV range, which is commonly referred to as the 'Compton hump'. (iii) If the inner disc is mildly ionized, we expect numerous emission lines at energies ${\le}2$\,keV. If they are relativistically smeared, then a smooth component should be observed, in excess to the power-law like continuum emission (e.g. \citealt{rossfabian05}). This feature could explain the ``soft-excess'' that is observed in many Seyfert galaxies. 

The ``negative'' X-ray time-lags (i.e. soft-band variations lagging the hard band variations observed at frequencies higher than $\sim 10^{-3}$ Hz) that have been detected in many Seyferts during the last few years have triggered a great deal of scientific interest on interpreting their nature. The first, tentative, detection was in AGN Ark~564 \citep{mchardy07}, where an origin in reflection from the accretion disc was proposed. The first statistically significant detection came from
1H~0707-495 \citep{fabian09}. Since then, such X-ray time-delays have been detected in many AGN (e.g. \citealt{emmanoulopoulos11,demarco13,kara16}). Positive X-ray time-delays (i.e. hard-band variations lagging the soft band variations, observed mainly at lower frequencies), have also been observed for quite some time in both AGN (e.g. \citealt{papadakis01,mchardy04,arevalo06}) and X-ray binaries (e.g. \citealt{miyamoto89,nowak96,nowak99}).

So far, the main result from the study of the negative, high-frequency time-lags is that their amplitude is small. This implies a short distance between the X-ray source and the reflective
medium. In many cases, the delays are so small that, if they depend on distance only, the X-ray source should be located at just a few gravitational radii (${\rm r}_{\rm g}={\rm GM}/{\rm c}^{2}$) on top of a disc that rotates around a rapidly spinning BH. However, although the absolute amplitude of the reverberation time-lags does depend on the height of the X-ray source (for a given BH mass), the {\it observed} time-lag amplitude also depends on ``dilution'' effects (i.e. the amount of the reflection component that is present on both energy bands that are cross-correlated) as well as on the amplitude of the continuum hard band time-lags. Definite conclusions regarding the geometry of the inner region in AGN (i.e. the X-ray source height and inner radius of the accretion disc) can only result from proper model fitting of the observed time-lag spectra.

Theoretical modeling of the observed time-lag spectra, taking into account relativistic effects, has been performed a few times so far. \citet{cackett14} modeled the time-lags between the $5-6\,\mathrm{keV}$ and $2-3\,\mathrm{keV}$ 
bands in NGC 4151. They assumed the lamp-post geometry (where the point-like source of irradiating hard X-ray photons is located at a height, $h$, above the BH; first proposed by \citealt{martocchia96}). They considered only a slowly rotating 
and a rapidly spinning Kerr BH (spin parameter a=0 and a = 0.998, respectively), and they studied the response of the 6.4 keV photons only. \citet[][; E14 hereafter]{emmanoulopoulos14} performed a systematic model fitting of the time-lags 
between the $0.3-1$ and $1.5-4\,\mathrm{keV}$ bands for 12 AGN. They also assumed the lamp-post geometry, and the calculated time-lags taking into account the full reflection spectrum (in the case of a neutral disc). \citet[][; EP16a hereafter]{epitropakis16a}
used a similar model to study the Fe-K$_{\alpha}$ time-lags (as a function of frequency) of a sample of AGN. They applied a new method to estimate the time-lags \citep[][; EP16b hereafter]{epitropakis16b}, and included all dilution 
effects. Recently, \citet{chainakun16} performed a simultaneous fitting of the time-averaged spectrum and the time-lags/energy spectrum in Mrk 335, IRAS 13224-3809 and Ark 564, using {\it XMM-Newton} data. They computed the time-lags 
considering the full reflection spectrum in the case of an ionized disc, including dilution effects.

In this work, we study the time-lags between 0.3--1 and 1--10 keV (as a function of frequency) of three AGN, namely Ark~564, MCG-6-30-15 and 1H~0707-495. These are X-ray bright and highly variable sources, and are among the most frequently observed AGN 
by  {\it XMM-Newton} (net exposure $\ge 600$\,ksec, for all objects). We focus exclusively on time-lags, although the variability amplitude (i.e. the modulus of the cross-spectrum) can also be very 
constraining (see e.g. \citealt{mastroserio18}). Ideally, we should model both the lags and amplitude, as well as the time-averaged spectrum, but this is a much more complicated work, which we leave for the future. We used the method of EP16b to compute 
the time-lags. The resulting time-lag spectra are of the highest possible quality, in terms of signal-to-noise ratio, they are 
unbiased, with known errors, and can be used to fit models to the data following traditional, ${\chi}^{2}$-minimization techniques. We then applied a new model, namely {\tt KYNREFREV} \citep{dovciak18}. The model estimates the 
theoretical time-lags between any two energy bands (from 0.1 up to 100\,keV) as a function of frequency and the time-lags (at a constant frequency) as a function of energy. The model assumes the lamp-post geometry, it takes into account all the 
relativistic and dilution effects, as well as the disc ionization. Moreover, the model is the first of its kind, as it can be used within {\tt XSPEC} \citep{arnaud96}\footnote{The model code is publicly available 
at https://projects.asu.cas.cz/stronggravity/kynreverb/. When in the site, go to Readme, for a detailed model description, and extensive instructions to download, use the code etc.}, to fit the observed time-lag spectra in the same way that other models are used in {\tt XSPEC} to fit the X-ray energy spectra.  

In Sec.~\ref{sec:sample} we describe the data and the sample and in Sec.~\ref{sec:lags} the time-lag estimation. In Sec.~\ref{sec:analysis} we show a brief description of the data reduction and subsequent analysis
with {\tt XSPEC}. Furthermore, Sec.~\ref{sec:analysis} provides
the results obtained using the model. Finally, in Sec.~\ref{sec:discussion} we discuss our results and give some clues into future developments of the code.

\section{Data analysis}  \label{sec:sample}

We chose Ark~564, MCG-6-30-15 and 1H~0707-495 because they are X-ray bright and highly variable AGN. They have been observed repeatedly with {\it XMM-Newton}, totalling 0.6, 0.7 and 1.3\,Msec of net exposure, respectively (see EP16a). We focused on the data provided by the {\it XMM-Newton} satellite because of its high sensitivity, large collecting area and high bandpass ($0.3-10$\,keV) of its detectors, as well as 
its ability to continuously observe an X-ray source for long periods of time (up to $40$\,h), which allows accurate determination of time-lags over a wide frequency range. We provide a summary of the X--ray properties of these sources in Sec.~\ref{sec:sample}.

The details of the {\it XMM-Newton} observations we used in this work are listed in Tab.~1 of \citet[][; EP17 hereafter]{epitropakis17}. The data reduction process is described in Sec.~2 of the same paper. Briefly, we considered EPIC$-pn$ data only, and we processed them with the Scientific Analysis System (SAS, v. 14.0.0; \citealt{gabriel04}). Source and background light curves were extracted from circular regions on the CCD, of a fixed radius of 800 pixels centred on the source coordinates listed on the NASA/IPAC Extragalactic Database. We extracted source and background light curves with a bin size of 10\,sec, using the SAS command {\it evselect}, in the 0.3--1 and 1--10 keV bands. We expect the presence of a strong reflection component in the former band in the case of X-ray reflection from ionized material. The higher energy band should be representative of the continuum component variations, mainly. It should also include the iron line photons as well, therefore we expect dilution effects to affect the observed time-lags. Nevertheless, we chose
to use this broad band in order to increase the signal-to-noise ratio of the time-lag estimates. In any case, our theoretical model takes into account dilution effects.  

Detailed description of the photon extraction from the events files, as well as the search for pile-up and the background subtraction can be found in EP17. There are a few missing points 
in the resulting light  curves, mostly randomly distributed throughout the duration of an observation. In some cases they appeared in groups of less than $\sim 10$ points. We replaced these points by a linear interpolation, with the addition of the appropriate Poisson noise. 

\subsection{The sample} 

{\bf ARK~564} ($z=0.02468$) shows the typical characteristics of a Narrow Line Seyfert I (NLS1): steep X-ray spectrum, a strong soft excess, and rapid variability. It also has the benefit of being extremely
bright in the soft X-ray band ($F_{0.3-10~\mathrm{keV}}=1.4\times10^{-10}$~erg/s/cm$^2$; \citealt{kara13b}). Because of these qualities, it has been observed by all the major
X-ray observatories, including {\it XMM-Newton}, {\it Suzaku} and {\it NuSTAR} \citep{kara17}. A low-ionization warm absorber has been detected in the X--ray spectrum of the source (e.g. \citealt{papadakis07,tombesi10,laha14,giustini15}). Previous analysis
of the X-ray continuum with {\it XMM-Newton} data have found photon indices of ${\Gamma}\sim 2.4  - 2.5$ \citep{papadakis07,demarco09}, also confirmed by {\it NuSTAR} \citep{kara17}.

Ark~564 was the first source where negative time-lags were, tentatively, detected \citep{mchardy07}. \citet{kara13b} measured its lag-energy spectrum at different temporal frequencies. They found that low-frequency lags ($\sim 10^{-5}-10^{-4}$~Hz) show a featureless log-linear increase with energy, while the high-frequency lags ($\sim 10^{-4}-10^{-3}$~Hz) showed clear evidence for iron~K reverberation. The soft-vs-continuum time-lags of this source have been fitted
by E14 and \citet{chainakun16}, while the iron line vs. the continuum time-lags have been fitted by EP16a. They found values for the height of the lamp-post
of $h=4.6^{+0.9}_{-0.7},5.0^{+0.6}_{-0.1},>28\,{\rm r}_{\rm g}$ (E14,\citealt{chainakun16} and EP16a, respectively). For the spin of the BH, E14 estimated
a value of $a=0.05^{+0.45}$, \citet{chainakun16} froze it to $a=0.998$ and EP16a found that it could not be constrained. The inclination angle of the inner accretion disc was found to be
${\theta}_{0}=58_{-12},45{\pm}1$\,deg.

{\bf MCG~6-30-15} is a well-studied NLSI galaxy and it was the first source where a broad iron K emission line was detected \citep{tanaka95}. Since the initial discovery with
{\it ASCA}, the source has been observed several times with a number of instruments, and its X--ray spectrum has been well studied \citep{guainazzi99,wilms01,fabian03,vaughan04,brenneman06,miller07,miniutti07,turner07,marinucci14}. It
shows clear evidence for warm absorbers \citep{otani96,lee01,turner03,turner04}. Recent studies, using simultaneous {\it XMM-Newton} and {\it NuSTAR} data, have shown that the spectral slope of the continuum
power-law is ${\Gamma}\sim 2.1$ \citep{marinucci14}.

The X-ray time-lags in this source were originally studied using the first 300~ks observation by \citet{emmanoulopoulos11}, who found a significant soft-lag at high frequencies and a hard-lag
at low frequencies. Modeling of the soft and the iron-line vs. the continuum lags have revealed a height of the lamp-post to be
of $h=2.9^{+0.4}_{-0.7},<3\,{\rm r}_{\rm g}$ (E14 and EP16a, respectively) and a BH spin of ${\rm a}=0.98_{-0.26}$ (E14). The inclination angle of the inner accretion disc was found to be
${\theta}_{0}=35_{-8}^{+11}$\,deg. (E14). \citet{kara14} found evidence of time-lags changing within two {\it XMM-Newton} observations, which might indicate a varying corona.

{\bf 1H~0707-495} ($z=0.0411$) shows a dramatic drop in flux at 7~keV in its time-integrated spectrum \citep{boller02}. This feature suggests the presence of
relativistic reflection effects \citep{fabian04}. \citet{fabian09} found from a 500~ks {\em XMM-Newton} that the residuals to a phenomenological continuum model contain both the broadened iron~K
and iron~L emission lines. \citet{mizumoto14} proposes another model to interpret these features based on the partial covering model. Previous analysis of the X-ray continuum with data 
from {\it XMM-Newton} provide photon indices in the range of ${\Gamma}=2.5-2.9$ \citep{dauser12}. Optical reverberation 
studies have given a BH mass estimate of ${\rm M}_{\rm BH}=(5.2{\pm}3.2){\times}10^{6}\,{\rm M}_{\odot}$ \citep{pan16}.

\citet{fabian09} and \citet{zoghbi10} showed that the high-frequency variations in the 0.3--1~keV lag behind those in the 1--4~keV band by $\sim 30$~s. This was the first significant detection of negative time-lags
in AGN. These authors interpreted this short timescale lag as the reverberation time delay between the primary-emitting corona and the inner accretion disc. This short time delay would put the corona at a height
of $<10\,r_{\mathrm{g}}$ from the accretion disc. Later, a short timescale lag was also found in the iron~K band using the 1.3~Ms of archival {\em XMM-Newton} data on the source \citep{kara13a}. Model fits of the soft
and the iron-line vs. the continuum time-lag spectra have resulted in best-fit values of $h=2.4^{+0.6}_{-0.3},<20\,{\rm r}_{\rm g}$, for the height of the lamp-post (E14
and EP16a, respectively) and a BH spin of ${\rm a}=0.32^{+0.24}_{-0.22}$ (E14). The inclination angle of the inner accretion disc was found to be
${\theta}_{0}=27_{-5}^{+9}$\,deg. (E14). \citet{done16} propose the possibility that this AGN could have a low BH spin.

\section{Time lags estimation}  \label{sec:lags}

We computed the time-lag spectra of each source following the method of EP16b. First, we divided the available light curves in each energy band into $m$ segments of duration $T=20$\,ksec (the number of segments is listed 
in Tab.~2 of EP17). For a given pair of segments we used standard Fourier techniques to calculate the  so-called cross-periodogram at frequencies $\nu_p=p/N\Delta t$, where $p=1,2,\ldots, N/2$ ($N$, is the total number 
of points in each segment). Our final estimate for the cross-spectrum (CS), $\hat{C}_{x,y}(\nu_p)$, was the mean of the $m$ individual cross-periodograms
at each frequency. We did not average over neighbouring frequencies, as this can introduce a bias at low frequencies (EP16b)\footnote{We note that, apart from smoothing, time lags may also be biased due to light curve binning and the finite light curve duration (i.e. due to red noise effects). All these effects actually affect first the real and
imaginary parts of the complex cross spectrum, as we have shown in EP16b. We
suspect that the real and imaginary parts may be ``less" biased than time lags (after all, the
mean of the argument of a complex number is not necessarily equal to the argument of the means). However, it is not easy and/or straightforward to quantify the difference.}. We only considered frequencies lower than half of the Nyquist frequency, to avoid the effects of 
the light-curve binning on the time-lag estimates. 

Following standard practise, we used the following equations in order to compute the observed time-lags and their error:

\begin{equation} \label{eq1}
\hat{\tau}_{xy}(\nu_p)\equiv\frac{1}{2\pi\nu_p}\mathrm{arg}[\hat{C}_{xy}(\nu_p)]
\end{equation}

and,

\begin{equation} \label{eq2}
\hat{\sigma}_{\hat{\tau}}(\nu_p)\equiv\frac{1}{2\pi\nu_p}\frac{1}{\sqrt{2m}}\sqrt{\frac{1-\hat{\gamma}^2_{xy}(\nu_p)}{\hat{\gamma}^2_{xy}(\nu_p)}} \mbox{ . }
\end{equation}
\noindent The quantity $\hat{\gamma}^2_{xy}$ in the equation above is the estimate of the so-called coherence function, which is defined as:

\begin{equation} \label{eq3}
\hat{\gamma}^2_{xy}(\nu_p)\equiv\frac{|\hat{C}_{xy}(\nu_p)|^2}{\hat{P}_x(\nu_p)\hat{P}_y(\nu_p)} \mbox{ . }
\end{equation}

\noindent $\hat{P}_x(\nu_p)$ and $\hat{P}_y(\nu_p)$ are the traditional periodograms of the two light curves, which are also calculated by averaging over the same $m$ segments. EP16b showed that, in the presence of measurement 
errors, if the light curves are intrinsically coherent (i.e. the intrinsic coherence function is equal to unity at all frequencies), then the 
observed coherence should be well fitted by a function like the one defined by equation (25) in EP16b. We fitted the observed coherence functions by this equation, and we determined the frequency $\nu_{\rm crit}$ at which the 
observed coherence is equal to $1.2/(1+0.2m)$ (EP16b). According to these authors, at frequencies higher than $\nu_{\rm crit}$ eq. (\ref{eq2}) underestimates the error of the time-lag estimates, their distribution is uniform, and their mean value converges to zero, irrespective of the intrinsic time-lag spectrum. On the other hand, at frequencies lower than $\nu_{\rm crit}$, eq. (\ref{eq2}) is a reliable estimate of the time-lags error and their distribution is (approximately) Gaussian (as long as $m\ge 20$). 

The observed time-lag spectra for the three sources, estimated up to $\nu=\nu_{\rm crit}$, are shown in Fig.~\ref{fig:frequency}. The time-lags are estimated over a frequency range that is much broader than the frequency range of the ``iron-line vs. the continuum" time-lags of EP16a. This is mainly due to the larger signal-to-noise ratio of the light curves we use in this work. This fact allows the reliable time-lags estimation at 
higher frequencies. 

The low frequency time-lags are positive in all sources. They are indicative of the continuum, hard lags, that are routinely observed in all X--ray bright, compact sources. On the other hand, negative (i.e. soft band lags) are clearly visible 
at frequencies which are typically higher than $\sim$ a few times 10$^{-4}$ Hz, in all three sources. We note that we have kept the $y-$axis the same in all sources (Fig.~\ref{fig:frequency}), so that the amplitude of both the positive and negative time-lags can be 
directly compared between them. 

\section{Model fitting and results}  \label{sec:analysis}

\subsection{The model} \label{model}

The {\tt KYNREFREV} model \citep{dovciak18} estimates the theoretical time-lags in the case of the lamp post geometry. The model was developed with the main aim to be used within {\tt XSPEC}. In this way, it is straight-forward to fit the model to observed time-lag spectra. In addition, the model is reasonably fast. For example, the estimation of the theoretical time-lag spectrum for a given set of parameters takes a few seconds only in a portable computer. The main physical parameters of the model are:

\begin{enumerate}
\item{The BH mass, M$_{\rm 8}$ (in units of $10^8{\rm M}_{\odot}$).}
\item{The angular momentum ($-1\le{\rm a}\le1$)), i.e.  the BH spin in geometrized units.}
\item{The height, {\it h}, of the X-ray source (in units of ${\rm r}_{\rm g}={\rm GM}/{\rm c}^2$).}
\item{The viewing angle, ${\theta}_{0}$, of a distant observer with respect to the axis of symmetry of the disc, and}
\item{The observed primary X--ray flux in the 2-10\,keV band, L$_{2-10}$ (in units of the Eddington luminosity, ${\rm L}_{\rm Edd}$).}
\end{enumerate}

\noindent The disc is assumed to be geometrically thin, Keplerian, co-rotating or counter-rotating with the BH, with a radial extent ranging from the inner edge of the disc, which is equal to the innermost stable 
circular orbit, ${\rm r}_{\rm ISCO}$, up to the outer radius (${\rm r}_{\rm out}=10^{3}\,{\rm r}_{\rm g}$). The spin of the BH uniquely defines ${\rm r}_{\rm ISCO}$. When measured in geometrized units, the spin can attain any value between zero for a Schwarzschild BH (in which case ${\rm r}_{\rm in}=6\,{\rm r}_{\rm g}$), and either -1 or 1 in the case of a counter and co-rotating  disc around a rapidly spinning BH (in which case ${\rm r}_{\rm in}=9\,{\rm r_g}$ and $1\,{\rm r_g}$, respectively). The primary X-ray source (i.e. corona) is assumed to be a point-like region, located on 
the rotation axis at height {\it h} above the BH. It emits radiation with a spectrum of the form ${\rm F}_{\rm p}{\propto}{\rm E}^{-{\Gamma}}\mathrm{e}^{-{\rm E}/{\rm E}_{\rm c}}$, where ${\rm E_c}= 300$\,keV. The 
source emits isotropically, in its own frame.

The model takes into account all the relativistic effects in the propagation of light from the primary source to the disc. It estimates the incident flux on each disc element, as a function of time, and then uses the {\tt REFLIONX}\footnote{https://heasarc.gsfc.nasa.gov/xanadu/xspec/newmodels.html} FITS table files \citep{rossfabian05} to compute the local reprocessing of the incident radiation in the ionized accretion disc.

The ionization of each disc element, $\xi(r,{\phi},t)$, is set by the amount of the incident primary flux, $F(r,{\phi},t)$, and by the density of the accretion disc, $n(r)$, as ${\xi(r,{\phi},t)}{\propto} F(r,{\phi},t)/n(r)$. Where ${\phi}$ is the azimuth. We 
assumed a constant-density disc configuration. This choice should not affect significantly the results (but see Sec.~\ref{fits:results}), because the radial dependence of any realistic density profile is much less significant than the radial
decrease of the disc illumination by the lamp-post corona (see, e.g., Fig.~3 by \citealt{svoboda12}).

The model also takes into account all relativistic effects in the light path from the disc (and from the primary source) to the observer. It estimates the disc ``response function'' (which determines the disc 
emissivity, as a function of time, at any given energy band) and its Fourier transform (i. e. the disc ``transfer function''). Using this function, and taking into account the amount of the primary and reflected
component fluxes in each energy band (to account for dilution effects), the code computes the model time-lags.

The model also includes a phenomenological, power-law like prescription for the hard (i.e. positive) time-lags at low frequencies, of the form:

\begin{equation}
\tau(\nu)={\rm A}\nu^{-s}.
\end{equation}

The power-law like assumption of the low-frequency positive lags is a good description of the observed time-lags in various energy bands of many AGN (e.g. EP17). The overall time-lags are equal to the 
sum of the reverberation model time-lags plus the hard time-lags model, as defined above. This prescription (i.e. the observed time-lags are equal to the summation of the two components) is valid as long as the geometry of 
the X--ray source/inner accretion disc stays the same, as explained by EP17.

\begin{table}
 \centering
 \begin{minipage}{75mm}  \caption{Best-fit parameters when we fit data with spin fixed to rapid and zero (left and right column, respectively). In all cases, ${\rm L}_{2-10}=0.01$\,(${\rm L}_{\rm Edd}$).}
  \label{log_results}
  \begin{tabular}{@{}lcc@{}}
  \hline
   $a$ \,(GM/c)                                     &      $0.99^{(f)}$                           &    $0.0^{(f)}$        \\
\hline
\hline
               &           ARK~564                                &              \\
 & (${\rm M}_{\rm 8}=0.023^{(f,a)}$, ${\Gamma}=2.5\,^{(f)}$)   &                          \\
\hline
   ${\theta}_{0}$ \,(deg.)                          &       ${\le}14$                        &   ${\le}60$                        \\
   $h$ (${\rm r_g}$)                                &       $2.8_{-0.5}^{+0.4}$                   &   $2.2_{-0.1}^{+0.4}$                           \\
   Density\,($10^{15}\,{\rm cm}^{-3}$)              &       $0.20_{-0.12}^{+0.23}$                &   $12_{-10}^{+7} $                             \\
   ${\rm A}$                                        &   $(8.4{\pm}1.2){\times}10^{-5}\,^{(t)}$    &   --                                            \\
   ${\rm s}$                                        &      $1.60{\pm}0.03^{(t)}$                  &   --                                            \\
   ${\chi}^{2}/{\nu}$                               &       $1.42\,(70/49)$                       &   $1.40\,(69/49)$                       \\
   p-value                                          &       $0.027$                               &   $0.03$                       \\
\hline
\hline
               &           MCG~6-30-15                            &                        \\
   & (${\rm M_8}=0.016\,^{(f,b)}$,  ${\Gamma}=2.0\,^{(f)}$)   &               \\
\hline
   ${\theta}_{0}$ \,(deg.)                          &      ${\le}30$                              &     ${\le}60$                             \\
   $h$ (${\rm r_g}$)                                &      $3.8{\pm}0.5$                          &   $3.4{\pm}0.8$                          \\
   Density\,($10^{15}\,{\rm cm}^{-3}$)              &      $170{\pm}150$                          &   $11_{-9}^{+11}$                  \\
   ${\rm A}$                                        &    $(2.7{\pm}0.4){\times}10^{-3}\,^{(t)}$   &   --                                      \\
   ${\rm s}$                                        &      $1.25{\pm}0.03$                        &   --                                            \\
   ${\chi}^{2}/{\nu}$                               &     $1.37\,(49/36)$                         &   $1.34\,(48/36)$                        \\
   p-value                                          &      $0.07$                                 &   $0.08$                       \\
\hline
\hline
               &           1H~0707-495                            &             \\
      &   (${\rm M_8}=0.023\,^{(f,c)}$, $\Gamma=2.5\,^{(f)}$)                                       &                      \\
\hline
   ${\theta}_{0}$ \,(deg.)                          &    $22_{-5}^{+4}$                           &   $21_{-11}^{+9}$                                  \\
   $h$ (${\rm r_g}$)                                &       $3.6{\pm}0.3$                         &   $3.1{\pm}0.3$                   \\
   Density\,($10^{15}\,{\rm cm}^{-3}$)              &    $190_{-40}^{+50}$                        &    $3.8_{-1.3}^{+2.1}$                               \\
   ${\rm A}$                                        &    $(4.1{\pm}0.4){\times}10^{-4}\,^{(t)}$   &   --                                            \\
   ${\rm s}$                                        &      $1.50{\pm}0.03^{(t)}$                  &   --                                            \\
   ${\chi}^{2}/{\nu}$                               &       $1.42\,(84/59)$                       &   $1.53\,(90/59)$                       \\
   p-value                                          &       $0.018$                               &   $0.005$                       \\
\hline
\hline
\end{tabular} 
\\
\\
Frozen and tied parameters in the fit are indicated
by {\it (f)} and {\it (t)}, respectively. Errors correspond to the 68\,\% confidence intervals. References for the BH mass estimates are as follows: (a) \citet{vestergaard06}, (b) \citet{bentz16}, (c) \citet{zhou05}.
\end{minipage}
\end{table}

\begin{figure}
\centering
 \includegraphics[bb=39 22 564 720,width=6.0cm,angle=270,clip]{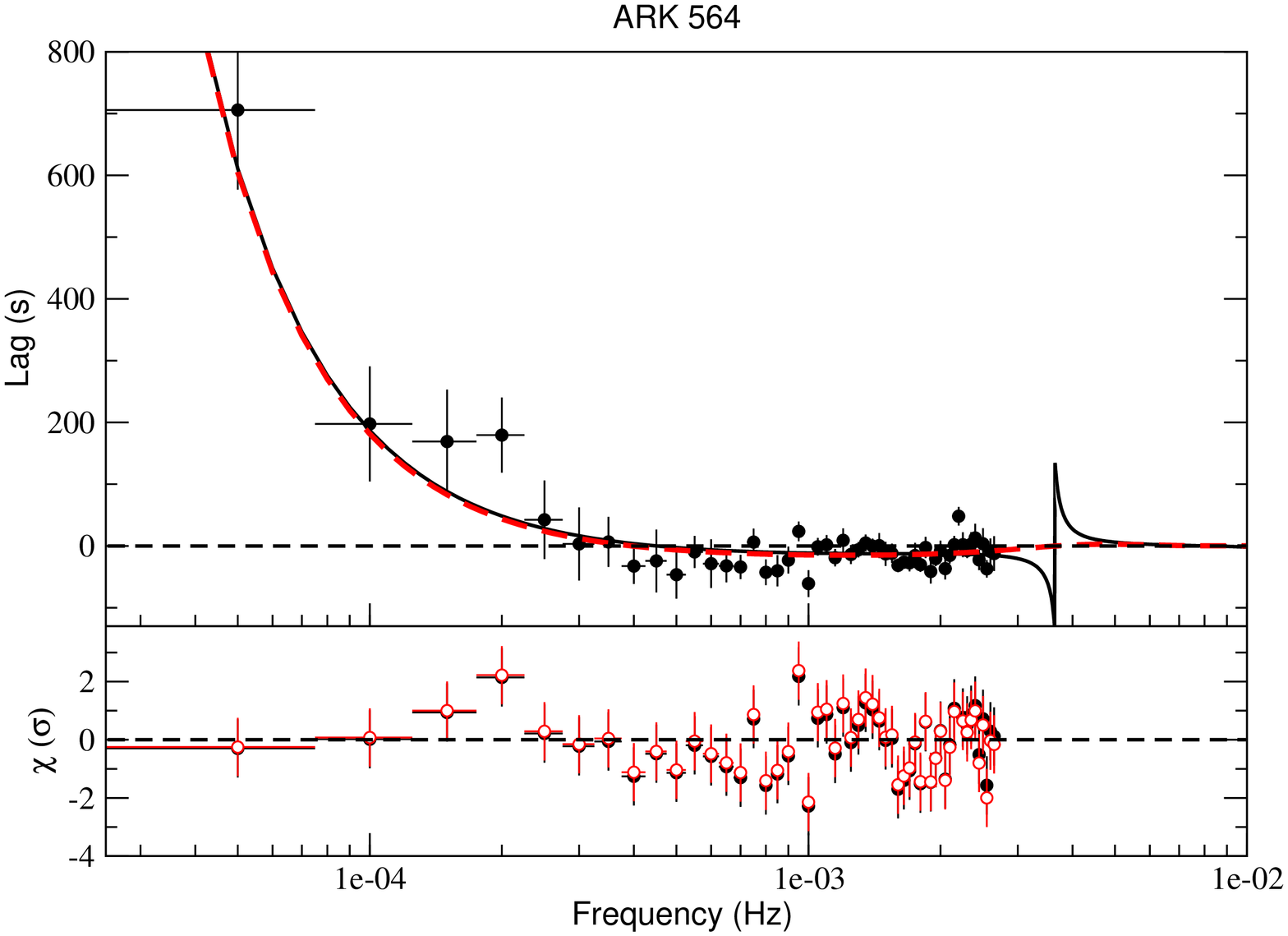}
 \includegraphics[bb=39 22 564 720,width=6.0cm,angle=270,clip]{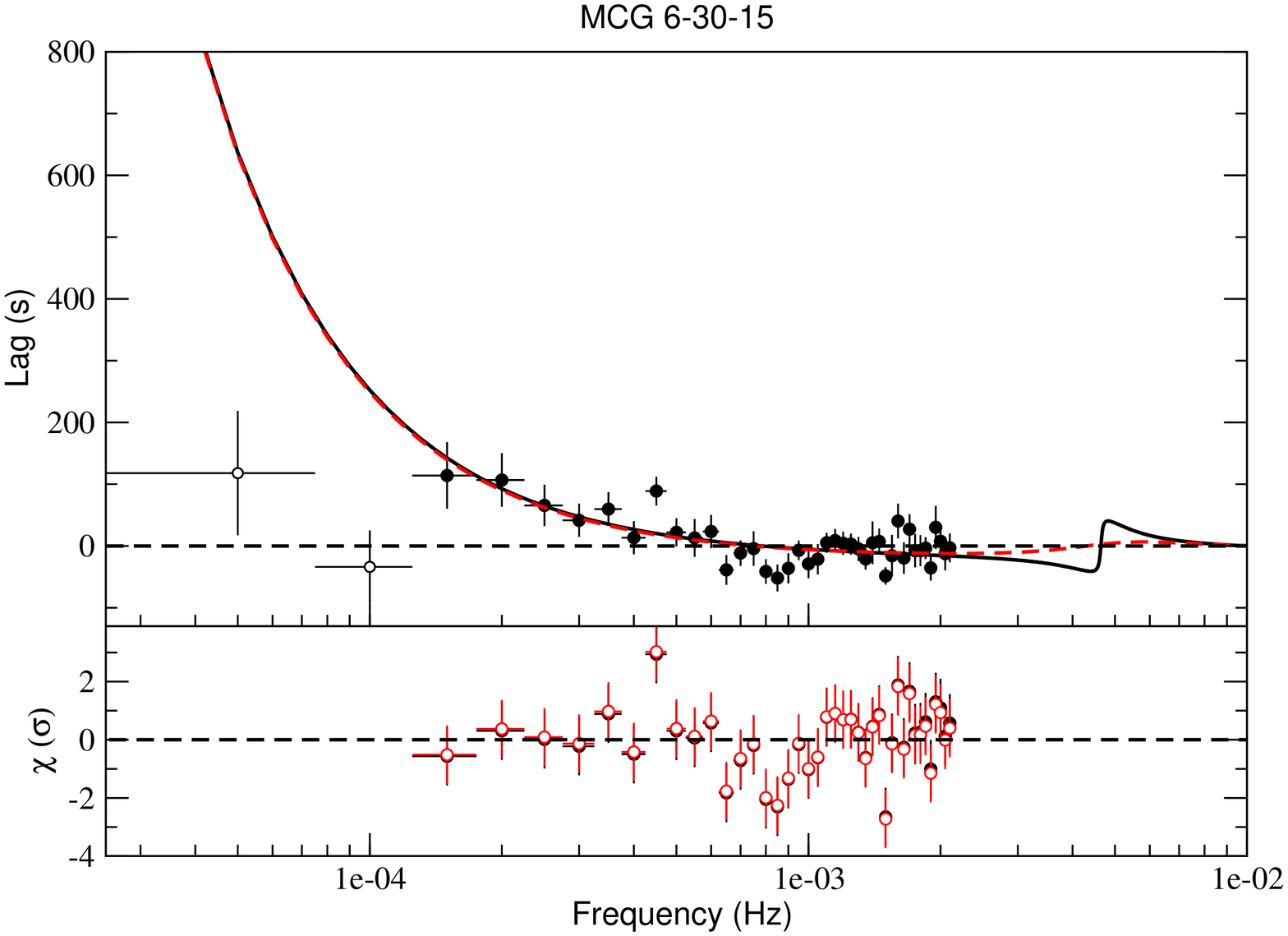}
 \includegraphics[bb=39 22 564 720,width=6.0cm,angle=270,clip]{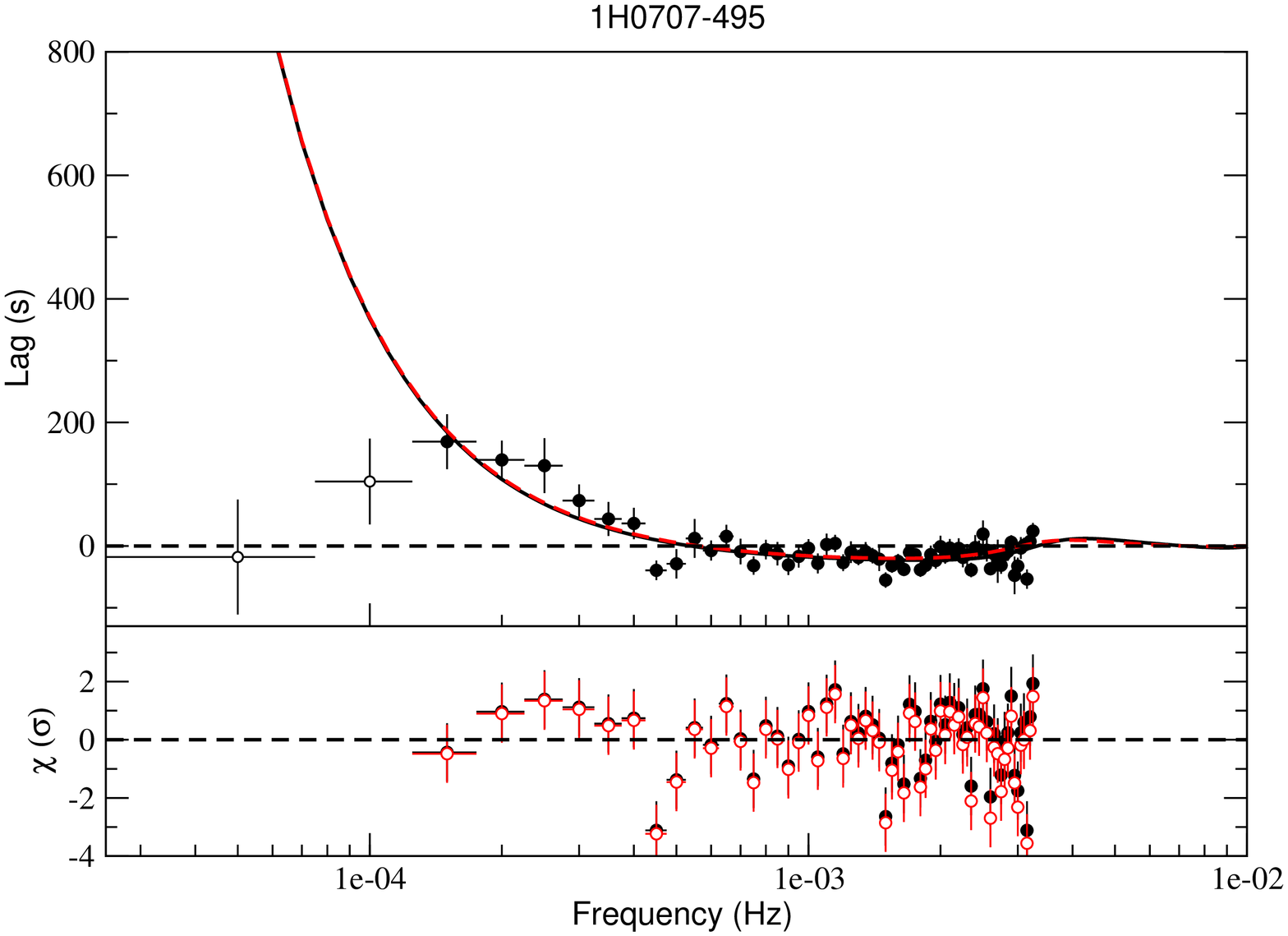}
\caption{The (0.3-1 versus 1-10\,{\rm keV}) X-ray soft time-lag versus frequency spectra of the three Seyferts. Solid black and dashed-red lines show the $a=0.99$ and $a=0$ fits of the data with the {\tt KYNREFREV} model (we used the filled 
points only for model fitting in the case of the data from MCG~6-30-15 and 1H~0707-495 -- see text for details). Filled black and open red circles indicate the $a=0.99$ and $a=0$ best-fit residuals. }
\label{fig:frequency}
\end{figure}

\subsection{The best-fit results}  \label{fits:results}

Although the time-lag spectra shown in Fig.~\ref{fig:frequency} are among the best time-lag spectra that can be estimated for AGN with the present day data, they are still not good enough to determine, accurately, all
the model physical parameters. For that reason, during the model fitting procedure we kept several of the model parameters fixed to certain values, as we explain below.

1) We considered only two BH spin parameters: $a=0$ and $a=0.99$. For a given BH mass and X--ray source height, the effects of the BH spin to the time-lag spectra are rather subtle (see e.g. EP16a and \citealt{dovciak14}). For that reason, we decided to fit the data in the two extreme cases of a non-rotating and a rapidly rotating BH. Our objective here is to investigate whether the existing data can indicate, with a high confidence, whether the BHs in these sources are rotating or not, irrespective of whether we can accurately determine the BH spin.

2) We kept the BH mass values fixed at the values listed in Tab.~\ref{log_results}. In the lamp-post geometry, the BH mass and the height of the X--ray source affect the amplitude of the 
reverberation time-lags in the same way: the larger the BH mass, and the larger the height, the higher the (negative) time-lags amplitude will be at low frequencies. If the adopted BH mass estimate is larger/smaller than the real BH mass in these systems, then the best-fit source height should be decreased/increased by the same factor. 

In fact, it is even possible that the model will not be able to fit the data if the adopted BH mass is too large. This is actually the case with 1H 0707-495. The BH mass estimate listed in Tab.~\ref{log_results} is based on 
the empirical relation between BH mass, luminosity at 5100 \AA\, and the full width at half maximum of H$\beta$ \citep{kaspi00}. Recently, \citet{pan16} using a similar method \citep{vestergaard06} and more recent data, published a BH mass estimate which is $\sim 2.3$ times larger that the value listed in Tab.~\ref{log_results}. When we repeated the model fit with the BH mass fixed at the new estimate, the best-fit height was pegged at the lowest possible value allowed by the model (i.e. $h=2.1$ and $1.1\,{\rm r}_{\rm g}$, when $a=0$ and $a=0.99$, respectively) and the fit quality was very poor ($\chi^2{\approx}200/59$ for both 
spins). If indeed the BH mass in this source is larger than $\sim (2-3){\times}10^{6}\,{\rm M}_{\odot}$, then the lamp-post model cannot explain its reverberation time-lag spectra.

3) We fixed the power-law index of the primary flux energy-spectra to the values listed in Tab~\ref{log_results}. These are similar to the best-fit spectral slope values that have resulted from model fits to recent X--ray spectra 
of the sources (see Sec.~\ref{sec:sample}). We repeated the model fits by adopting $\Gamma=2$ for Ark~564 and 1H~0707-495 and $\Gamma=2.5$ in the case of MCG~6-30-15. The best-fit results were almost identical to those listed in Tab.~\ref{log_results}. This 
shows that the chosen $\Gamma$ values should not seriously affect either the quality of the fit or the best-fit parameters.

4) We also kept  ${\rm L_{2-10}}$ fixed at 1\% of the Eddington limit, in all sources. The value of this parameter affects the ionization state of the disc (for a given BH mass and $h$). We discuss in the next section how 
does this choice affect our results.

5) We kept the iron abundance fixed at one. If the iron abundance is significantly larger than one, as could be the case for 1H~0707-495 (see for example, \citealt{fabian12}), then the reflection component will be 
stronger, and the resulting best-fit height will increase. We generated model time-lag spectra, using the best-fit parameter values listed in Tab.~\ref{log_results} with iron abundance equal to 1 and 7, and we saw that 
they differ by an amount smaller than the error of the observed time-lags error. We therefore conclude that the use of an iron solar abundance should not affect significantly our results.

We observe a low frequency turn-over
in the time-lag spectra of 1H~0707-495 and MCG~6-30-15 (see Fig.~\ref{fig:frequency}), in agreement with the results of EP17. This turnover could be due to warm absorber variations which respond to changes in the
ionising source (e.g. \citealt{silva16}). In any case, we decided to ignore the time-lags at the two lowest frequencies in these two sources, because the best-fitting results change significantly depending on whether we keep them or not. 

The best-fit results are listed in Tab.~\ref{log_results}. We report the $1\sigma$ errors (and upper limits) on the best-fit model parameters. Red and black lines in Fig.~\ref{fig:frequency} show the best fit models in the case of a Schwarzschild and a 
rapidly rotating BH, respectively. Best-fit residual plots are also shown in Fig.~\ref{fig:frequency}.

We note that we fitted the observed time-lags with the $a=0$ and $a=0.99$ models simultaneously, and we kept the amplitude and index of the empirical power-law component (which accounts for the low-frequency hard lags) tied between
the two models during the fits. In this way, we force the two reverberation models to fit the data assuming the same hard lags component. Since we do not have a prior expectation of what the continuum lags should be, we also fitted the data with the $a=0$ and $a=0.99$ models by freeing the power-law parameters, i.e. by assuming different continuum lags to one another. Not surprisingly, the models fit the data as well as when we keep the power-law parameters tied. The best-fit power law parameters are consistent (within the errors) between the $a=0$ and $a=0.99$ models, and with the best-fit parameters listed in Tab.~\ref{log_results}. The best-fit inclination, height and disc density values are also consistent with the values listed in Tab.~\ref{log_results}, although their errors are $\sim 10-30$\% larger.

Our results suggest that the X-ray source in these sources should be located very close to the central BH. For example, according to our best-fit results, the X--ray source height may be as low as $\sim 2.5{\rm r_g}$ in Ark 564. We conclude that the observed time-lags in these Seyferts strongly suggest a compact X-ray source, which is located close to the BH, at a height smaller than $\sim 4 {\rm r_g}$. This result {\it does not depend on the BH spin.} No matter whether the BH is rapidly spinning or not,  the height of the X-ray source should be very low in these sources.  

There is no significant statistical difference between the quality of the best-fits in the rapidly and zero spinning BH scenarios in Ark 564 and MCG-6-30-15. Both models can fit equally well the data with different combinations of $h, {\theta}_{0}$, and disc density (i.e. ionization state). However, this is not the case in 1H 0707-495. The $p-$value of the $a=0$ best-fit is low. The best-fit residuals of the $a=0$ fit do not show a significantly different structure than the $a=0.99$ residuals (bottom panel in Fig.\,~\ref{fig:frequency}), but they are systematically worse, specially at high frequencies. We therefore conclude that the hypothesis of a non-rotating BH in 1H 0707-495 is not supported by our results.

Statistically speaking, all the other fits are formally accepted, with $p-$values larger than 1\%. However, in Ark~564 and 1H~0107-495, the $p-$values are quite low (around $2-3$\%). The poor quality 
of the model fits is mainly due to the wavy-residuals which are clearly present in the respective best-fit residuals. A similar pattern is less pronounced in MCG-6-30-15 (hence the better quality of the best-fits in this case). This result suggests that the X--ray source/disc geometry in the inner region of AGN may be more complicated than the lamp post geometry. 

\subsection{The effects of disc ionisation}

The disc ionization affects significantly the strength of the reflection component at low energies. Therefore, it should also affect the amplitude of the reverberation time-lags. In our model, the disc ionization is set by the amount of the incident primary flux and the density of the accretion disc. For a given BH mass and source height the incident flux on the disc is determined by ${\rm L_{2-10}}$. In order to investigate how the disc ionisation affects our results, we refitted 
the data assuming that: a) the disc density is the same at all radii and ${\rm L_{2-10}}= 10$\% of the Eddington limit, and b) the disc density decreases with increasing radius as $r^{-1}$ and $r^{-2}$ (while ${\rm L_{2-10}}= 0.01\,{\rm L_{Edd}}$, like the fits we performed so far).

Our results showed that the best-fit $h$ values remained small (i.e. less than $\sim 5-6 r_g$), in all cases and for all sources. Depending on the model fit, the other model parameters were changing in such a way that the best-fit source height was always low. Neither did the best-fit  statistics change, except in 1H~0707-495. When we increased L$_{\rm 2-10}$ to 0.1 the $a=0$ model provided a good fit to the data. This was also the case when we considered the variable disc density. Additionally, the quality of the $a=0.99$ best-fit model when $n(r)\propto r^{-2}$ was poor ($\chi^2=102$ for 59 degrees of freedom). If the BH is indeed rapidly spinning in this source, the disc density cannot be decreasing very fast with radius. 

\section{Discussion and conclusions} \label{sec:discussion}

We report the results obtained by fitting the {\tt KYNREFREV} model to the 0.3--1 versus 1--10\,keV time-lags in three X--ray bright AGN (namely, Ark~564, MCG~6-30-15 and 1H~0707-495). These are among the most frequently observed AGN 
with {\it XMM-Newton}. We fitted the data by assuming either a rapidly spinning or a non-rotating BH. {\tt KYNREFREV} can be used to fit both the energy and the time-lags (either plotted as a function of energy, at a given frequency, or as a function of frequency, between two energy bands). In fact, this should 
be the ideal approach. Simultaneous fitting of the time-lags and the average energy spectrum spectrum has the potential to break a lot of degeneracies in our model fit results (see below). It should be possible, for instance, to constrain 
the ionization parameter fairly well by also fitting the model to the spectrum. Nevertheless, it is not a straight forward exercise, specially for highly variable sources. We are working on this, but in this paper we present our results 
from the model fit to the average time-lag spectrum only. The main results can be summarized as follows:

1) The X--ray source is located very close to the central BH. The best-fit height values are less than $\sim 4$ r$_{\rm g}$. This is a major result of our study. It does not depend on the BH spin and/or the disc ionization but it does depend on the assumed BH mass. If the adopted M$_8$ value is close to reality, then the observed time-lag spectra  really suggest that the X-ray source in these objects is located very close to the central BH. 

2) Both the high and zero spin models can fit the data well. 

3) The best-fits to the Ark 564 and 1H~0707-495 data are of rather low quality. We detect residuals around the best-fit reverberation model time-lags at high frequencies. This result suggests that the simple lamp-post geometry does not fully explain the X-ray source/disc configuration in AGN. 

Our best-fit heights are different than those reported in the literature but it is not straight forward to compare our results with the previously published results (which we list in Sec.~\ref{sec:sample}). We can think of many reasons for these 
differences. For example, we and E14 use more or less the same data, and the same energy bands to estimate the time-lag spectra, but they assumed different BH masses, and used different time-lags estimation methods. This is the 
reason why the time-lags vs. frequency plots in our Fig.~\ref{fig:frequency} and in their Fig~8 are not the same, despite their overall similarity. Our time-lag spectra have a much better frequency resolution, different normalization
at low frequencies, and they cover a different frequency band. These differences can affect the best-fit results. In addition, E14 did not consider the disc ionization in their model. This is an important issue, which can affect the 
best-fit height values. In fact, it is worth mentioning here that we use {\tt REFLIONX} to estimate the disc ionization, which assumes a constant density slab. Instead, assuming vertical hydrostatic equilibrium can reduce the 
amount of soft emission in the reflection spectrum (e.g. \citealt{nayakshin00}). Therefore, if we had used a hydrostatic equilibrium model, the best-fit source heights could be smaller than the ones we report in this work.

\subsection{Limitations of our modeling} 

A caveat of our analysis is that we used all the combined data together in the calculation of the time-lags. It is possible that the time-lag spectra vary with time (if, for example, the X--source/disc geometry, i.e. height and inner disc radius, varies 
with time). For example, \citet{alston13} analysed the data from NGC~4051 (a source not included in our sample) and they found indication of differences between the time-lags in various flux states of the source. Indeed, we cannot divide the 
available data for the sources we study in various flux states, as the number of the resulting light curve segments will not be large enough to allow for a statistically significant comparison between the resulting 
time-lag spectra. We suspect it is not possible to investigate, with the current data sets, whether the time-lag spectra of the sources in our sample vary with time or not, but we plan to investigate this in a future work.

The estimation of the model time-lags will be discussed in detail in \citet{dovciak18}. Roughly speaking, the model estimation is done along the lines explained in Sec.~4 of EP16a. A crucial step in the model estimation is the 
determination of the so-called disc response function, which determines the response of the disc to an instantaneous flare of continuum emission. In doing so, the model in its current form does not take into account the response of EPIC-$pn$, and  
assumes that all the emitted photons in both bands are detected with the same probability by the instrument. Although this assumption may apply (to some extent) to the low energy band, this should not be the case for the 1--10\,keV band, where the 
response of the instrument varies significantly across this band. As a result, the average energy of the photons detected in this band will be softer than the energy in the case when the EPIC-pn response was energy independent. It is difficult 
to quantify the effects of this problem. The amplitude of the low-frequency, hard lags will be smaller than in the case of a flat response (as these time-lags depend on the energy separation between the light curves). As for the reverberation 
time-lags, as long as the continuum photons detected in the 1--10 keV band trace well the continuum variability, the time-lag spectrum should not be significantly affected. To test this hypothesis, we used the best-fit parameter values listed in 
Tab.~\ref{log_results} in the case of 1H~0707-495, and we calculated the model time-lags as a function of frequency for the energy bands we considered as well as the 0.3--1 vs. 1--4\,keV, and the 0.3--1 vs. 2--5\,keV time-lags. In this way we try to 
investigate the effects of the different mean energy in the reference energy band when we take into account the response of the detector. We found that the difference between these model time-lag spectra is smaller than the error of the observed 
time-lags. Therefore, we do not expect that the fact of not taking into account the response function of EPIC-$pn$ to significantly affect our results. This may affect the time-lags (at a certain frequency) vs. energy plots (depending on the energy 
limits of the reference band, on the width of the energy bands etc). In any case, we plan to include the response of the detector in future versions of the code.

As we explained in Sec.~\ref{model} we assumed a power-law model for the continuum, hard lags, and we then added this model to the reverberation time-lags. Whatever is the reason for the continuum time-lags, it may result in spectral slope variations 
of the continuum. We have modeled the reverberation time-lags assuming an X-ray source with a constant spectral slope and a variable amplitude, without taking into account the hard time-lags, intrinsic to the X--ray source. Recently, \citet{mastroserio18} 
explored a more self-consistent way of including the continuum time-lags when estimating the reverberation time-lags and showed that adding the continuum lags as a separate model component without accounting for the resulting non-linear effects can 
somewhat bias the results. We suspect that this effect should not significantly bias our results, at least when compared to the biases that are present from all the other effects, but it has to be addressed in a future work.

\subsection{The black hole spin}

Our work shows that model fitting of the time-lag spectra alone cannot tell whether the BH in these AGN is rapidly spinning or not rotating at all. This is in agreement with the results of EP16a who showed that, in the case of the lamp-post geometry, the BH spin does not affect significantly the disc transfer/response functions (for
a given BH mass and source height). Their work did not include the effects of disc ionization, but our modeling does take into account these effects. 

Including disc ionization to the model has two noticeable effects. First, the ionized disc results in a stronger reflection component, specially in the 0.3-1\,keV band. Consequently, the X-ray source height turns out to be significantly higher than when the disc ionization is absent. This is the case for example with 1H~0707-495. E14 fitted the soft band time-lags of this source using a model similar to ours but without considering the effects of ionization.  They found $h=2.4^{+0.6}_{-0.3}\,{\rm r}_{\rm g}$, and a moderate spin of $a\sim 0.3$. We find a significantly larger height ($h=3.6\pm 0.3$\,r$_{\rm g}$) when $a=0.99$. For small source heights, the X--ray
continuum is significantly reduced (due to light bending, most of the continuum flux
goes directly to the BH and not to the observer). Our best model fit prefers a
higher height, because we increase the soft excess (due to ionisation), and at the
same time, the increase of the continuum dilution (due to the larger height light bending is not as important) is not
as much as the increase in the soft excess due to ionisation.
Low height (i.e. $h< 3\,{\rm r}_{\rm g}$) and a rapidly spinning BH is a forbidden configuration for this source. Such a model would predicts such a high reflection fraction that phase-wrapping should appear, contrary to what we observe. 

Secondly, disc ionization is one more extra parameter that can be adjusted in such a way so that both the $a=0.99$ and $a=0$ models can fit the data equally well. In the $a=0$ models, the source height and the disc density usually adjust themselves in such a way so that the strength of the reflection component turns out to be the same as with the $a=0.99$ models, contrary to the smaller distance between the X-ray source and the disc inner radius in the former case ($a=0$).

The only way to investigate the BH spin with the time-lags is if we manage to improve the observed time-lag spectra. This does not necessarily mean smaller error bars, but rather extending the time-lags estimation to higher frequencies. The $x$--axis in Fig.\,\ref{fig:frequency} is extended up to $10^{-2}$ Hz. The main reason was to plot the best-fit models at high frequencies and show a peculiar "spike" in the $a=0.99$ best-fit model to the Ark 564 and MCG-6-30-15 data, at around $\sim (3-5)\times 10^{-3}$ Hz. 

This feature is due to the so-called phase-wrapping: since the phase-lags are defined on the interval ($-\pi, \pi$], it is possible that this function will exceed $\pi$  at a certain frequency. In this case, it will ``flip back" to the value of $-\pi$ (and vice-versa). The frequency at which phase wrapping will occur depends on the strength of the reflection component. The stronger this component in the 0.3--1 keV band, the lower the frequency at which the phase wrapping will occur. In fact, this frequency is low enough in the case of 1H~0707-495 to discard the rapidly spinning BH/low lamp height solution. So, based on the best-fit model plots in Fig.\,\ref{fig:frequency}, if we manage to extend the Ark 564 and MCG-6-30-15 time-lags estimation at least to $\sim (5-6)\times 10^{-3}$ Hz,  we will be able to tell whether the BH is rapidly spinning or not. 

This is in particular the case for Ark~564. Both the spin zero and rapidly spinning solutions for this source suggest a low-height lamp and an almost face-on disc geometry, but the difference in the disc ionization is significant. For the rapidly spinning
solution the ionization is much higher than the case of the zero spin BH. This translates into a dramatic phase-wrapping effect at high frequencies, because of the extreme reflection for the case of rapidly spinning BH.

\begin{figure*}
\centering
 \includegraphics[bb=39 22 564 720,width=6.0cm,angle=270,clip]{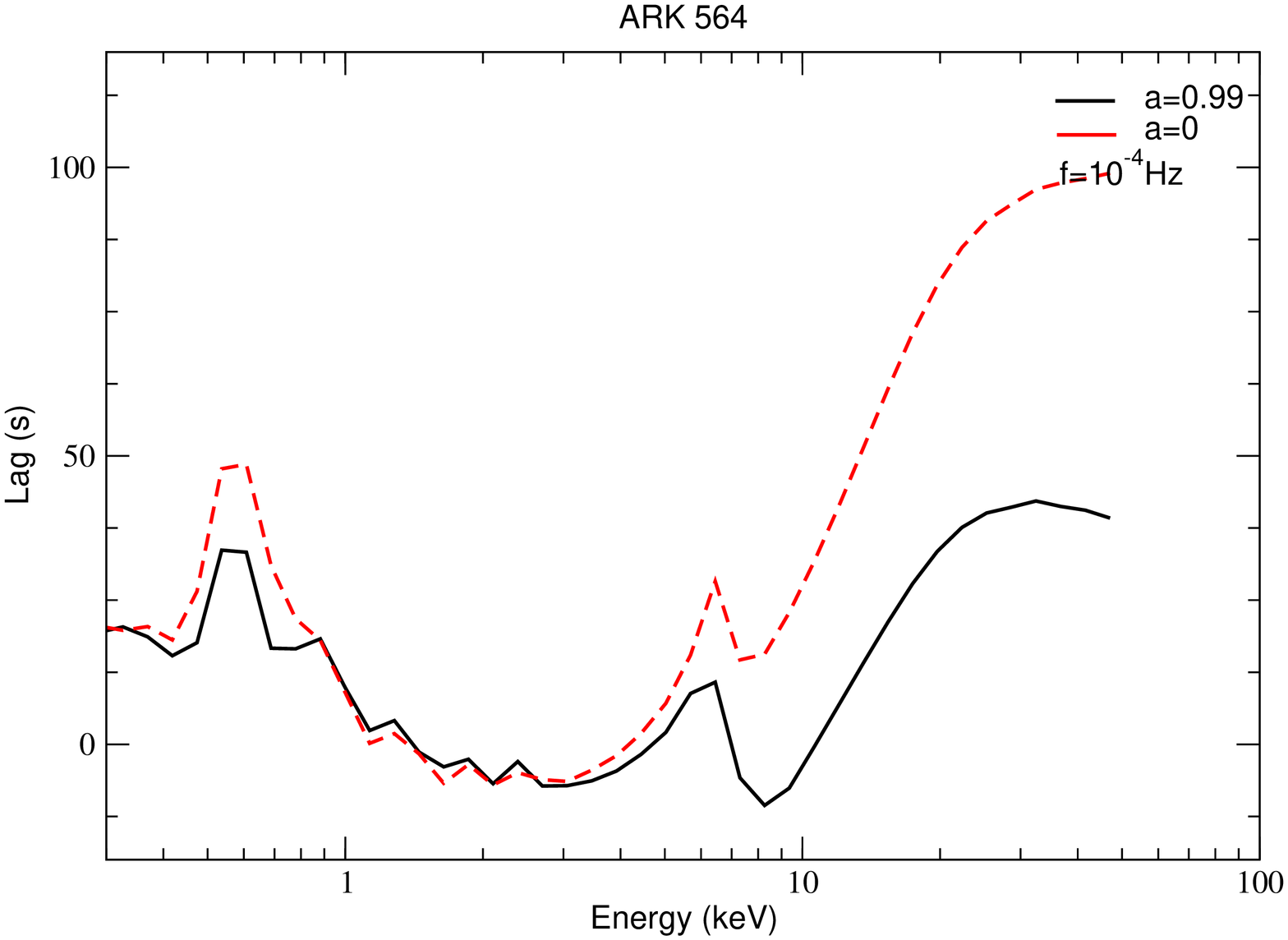}
 \includegraphics[bb=39 22 564 720,width=6.0cm,angle=270,clip]{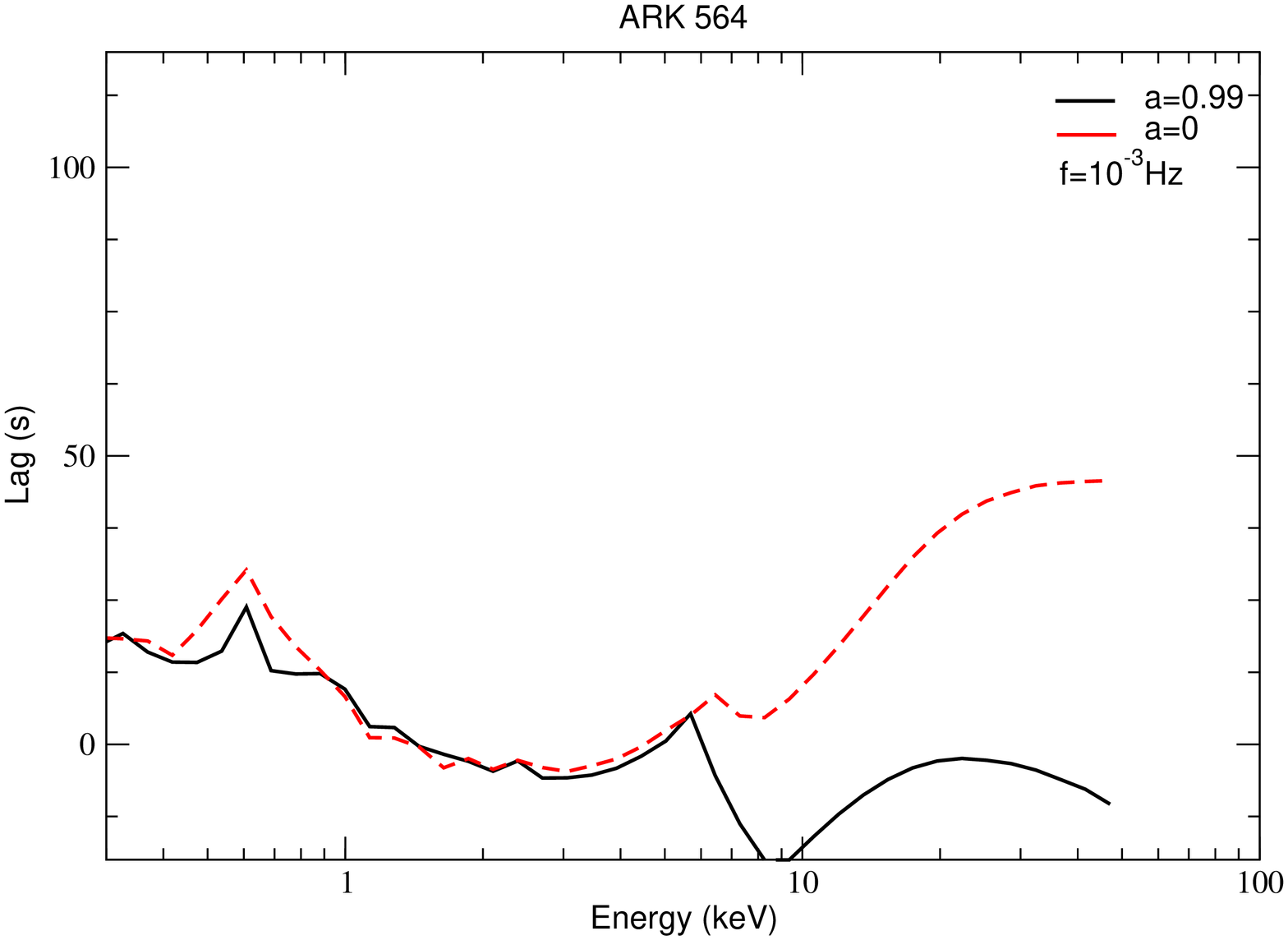}
 \includegraphics[bb=39 22 564 720,width=6.0cm,angle=270,clip]{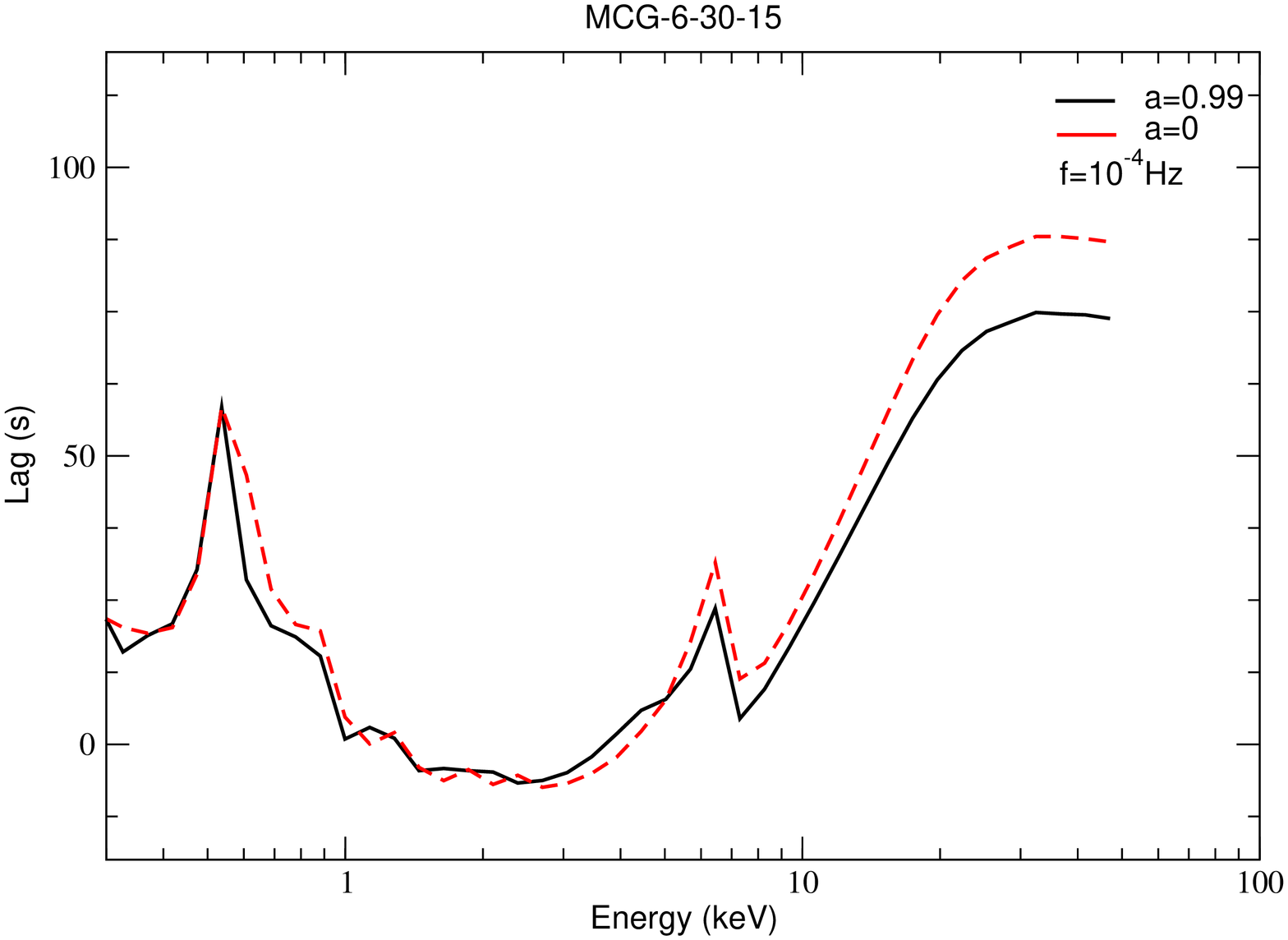}
 \includegraphics[bb=39 22 564 720,width=6.0cm,angle=270,clip]{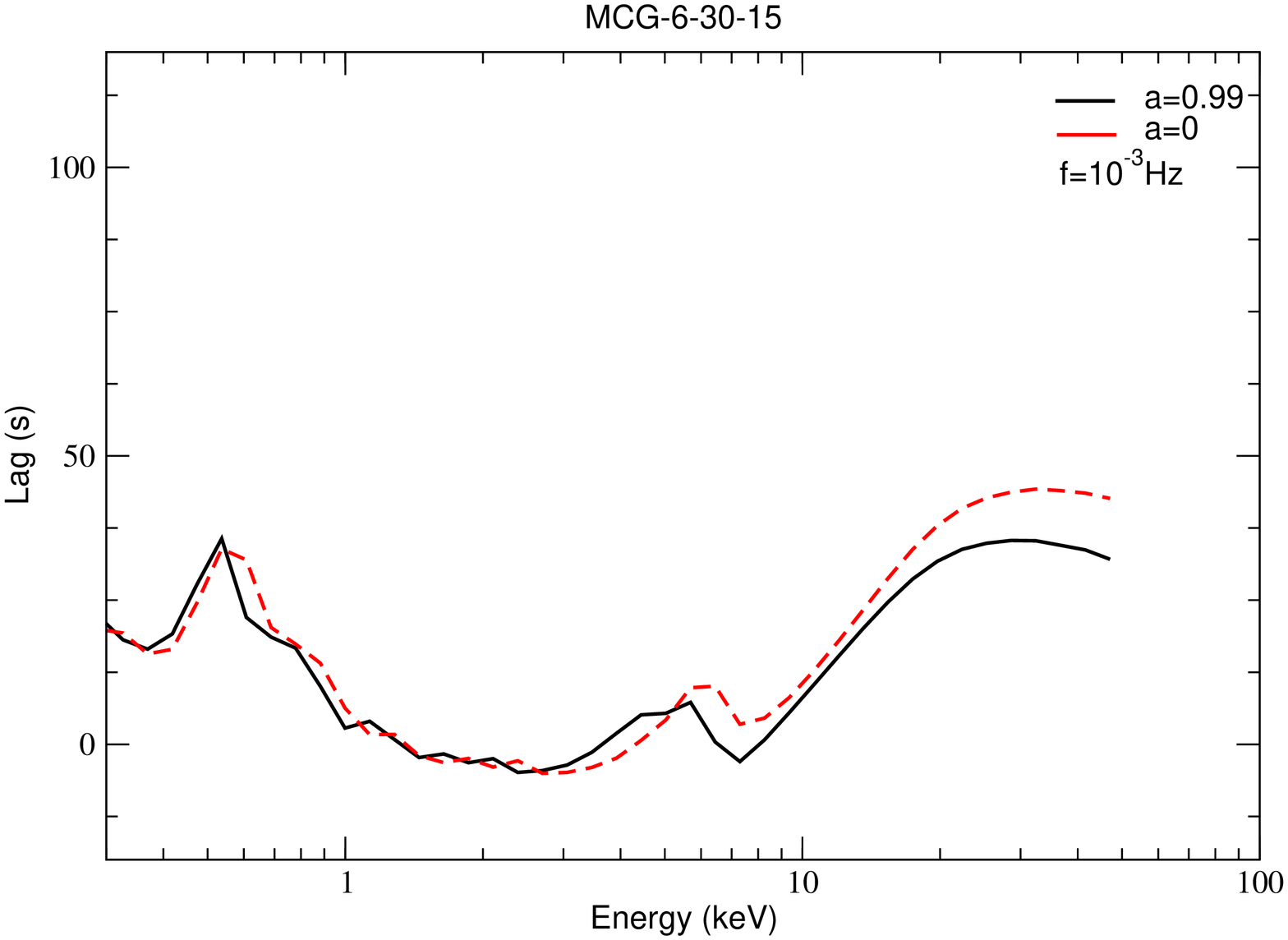}
 \includegraphics[bb=39 22 564 720,width=6.0cm,angle=270,clip]{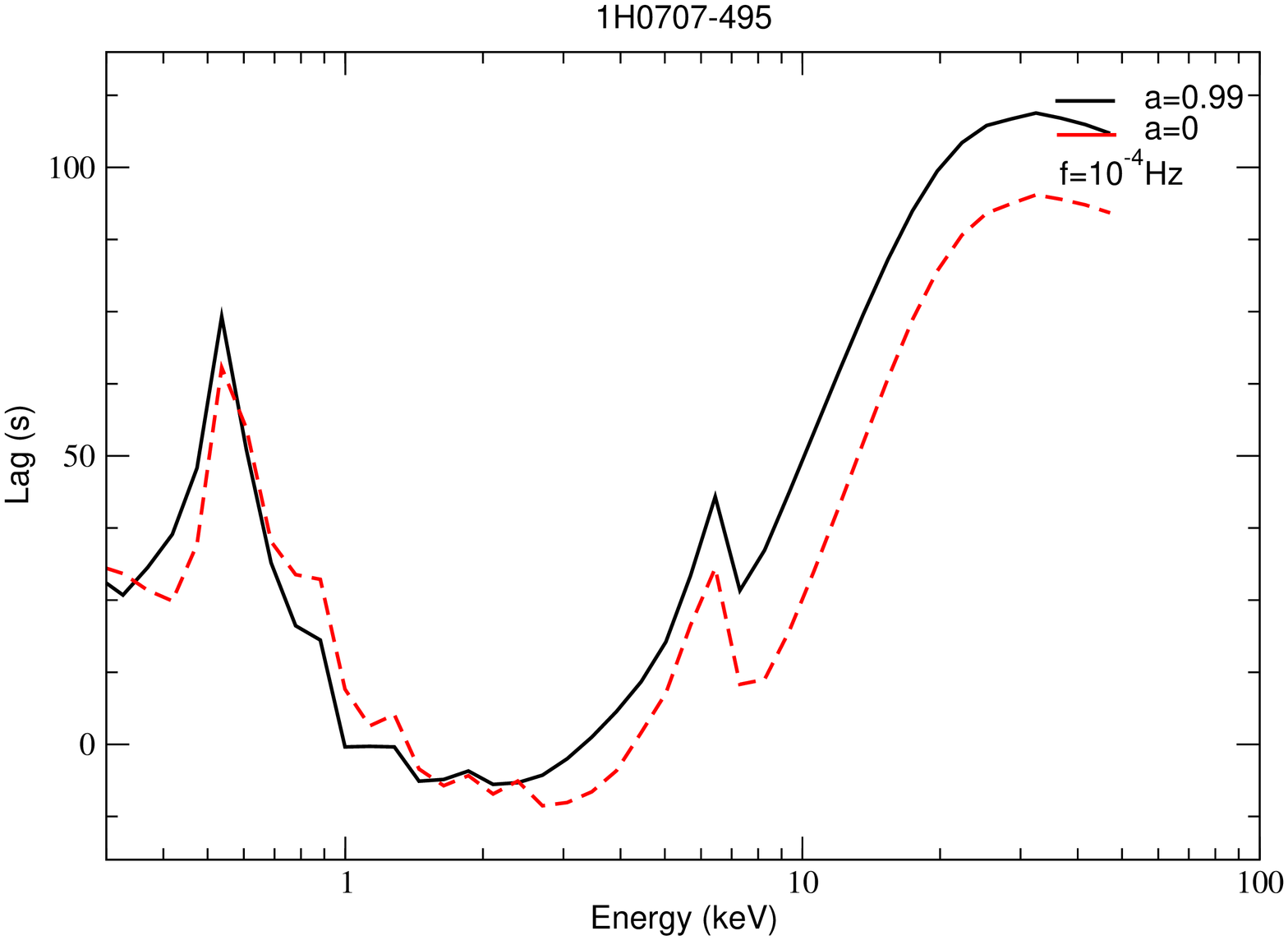}
 \includegraphics[bb=39 22 564 720,width=6.0cm,angle=270,clip]{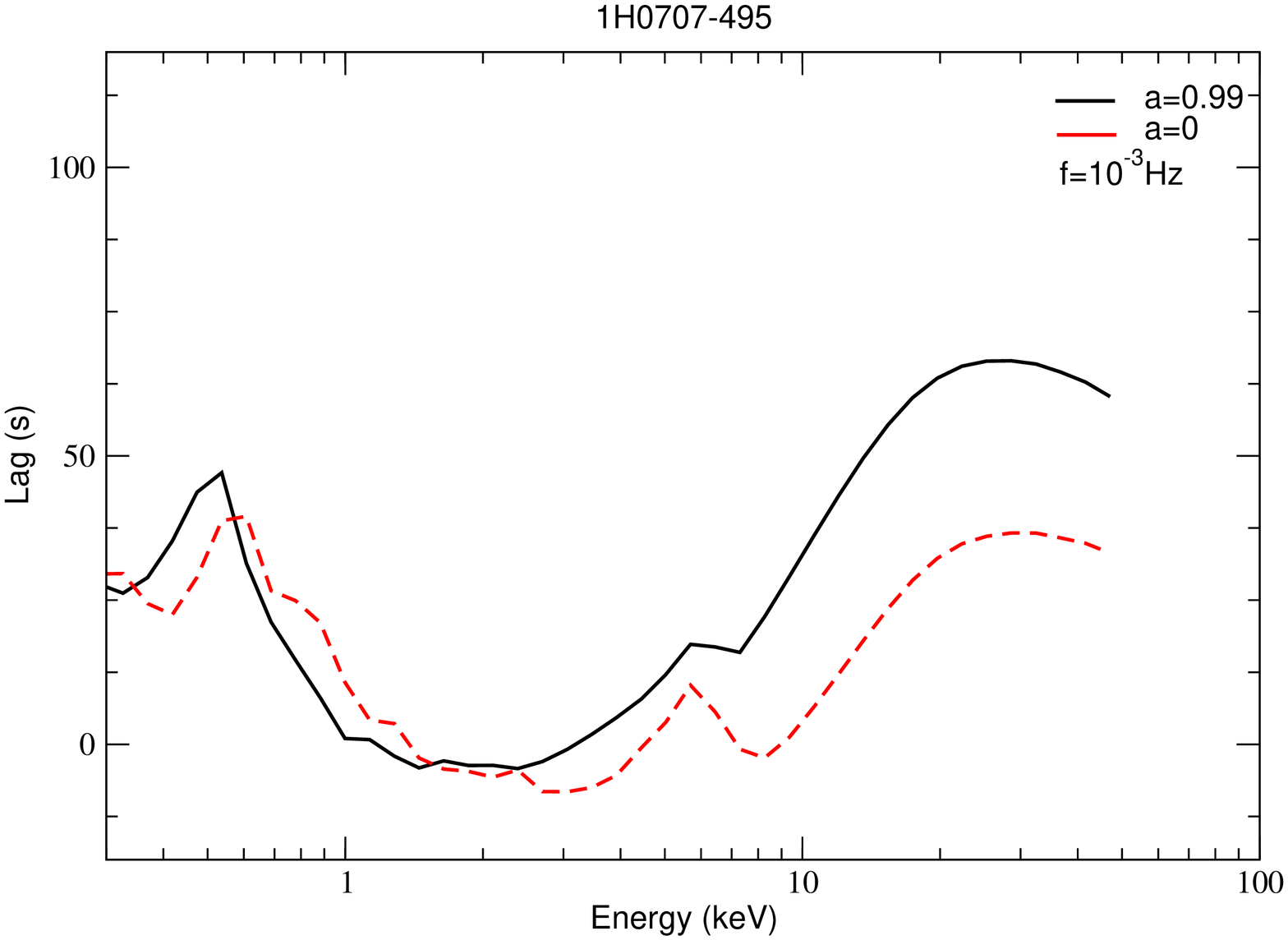}
\caption{The X-ray time-lags versus energy spectra (with reference energy band at 1-10\,keV) at $10^{-4}$\,Hz and $10^{-3}$\,Hz centroid frequencies (left and right, respectively) of the Seyfert-I AGN reported in this work. They are the result of the best-fitted {\tt KYNREFREV} models (solid-black and dashed-red correspond to rapid and zero spin, respectively) of the time-lag versus frequency spectra (Fig.~\ref{fig:frequency}).}
\label{fig:spectra}
\end{figure*}

Another possibility to investigate the BH spin in these objects is to study the time-lags at a fixed frequency as a function of energy. Fig.~\ref{fig:spectra} shows the predicted time-lag spectra versus energy from our best-fit models. The left 
and right panels show the $a=0.99$ and $a=0$ model time-lags (black and red lines, respectively) at 10$^{-4}$ and 10$^{-3}$ Hz frequencies (left and right panels, respectively). In all cases, time-lags are measured with respect the 1--10 keV band. These 
plots show that the largest differences between the high and zero spin solutions appear at energies above 10 keV. 

The time-lags vs. energy plots are almost identical in the case of MCG-6-30-15. There are differences at energies above $\sim 10$ keV in 1H~0707-495 and, especially, in Ark~564. Both the zero and rapidly spin solutions for this source suggest a low-height lamp and an almost face-on disc geometry. However, the best-fit height is smaller in the Schwarzschild solution. The difference between the $a=0$ and $a=0.99$ time-lags energy spectra above 10 keV could be due to the fact that the X--ray source in the $a=0$ case should be located  just above the horizon. As a result, the strength of the reflection component (which originates much further away) will be much more pronounced in this case.

However, it is not straight-forward to estimate time-lags at energies above 10 keV with the current data sets. The only usable data are those collected by {\it NuSTAR}, however the resulting light curves are not evenly sampled, and their length may not  be long enough for a reliable estimation of the time-lags at these energies.

\subsection{The case of an extended corona}

As we mentioned in Sec.~\ref{sec:analysis} the time-lag versus frequency spectra show wavy-residuals which are clearly present in the best-fit residuals, in particular in 
the case of Ark~564 and 1H~0707-495 (see the residual plots in Fig.~\ref{fig:frequency}). These residuals are responsible for the relatively poor statistics of the best-fit models, and suggest that the simple lamp-post geometry (which the {\tt KYNREFREV} model assumes) may not be an accurate representation of the X--ray/disc configuration in the inner region of AGN. 

This may not be surprising since the lamp-post geometry, although widely assumed in the literature, is probably a simplification of the real situation. A more complicated model has been recently proposed by \citet{chainakun17}. Their model consists of 
two co-axial point sources illuminating the accretion disc. It assumes that the variations of two X-ray sources are triggered by the same primary variation, but the two sources respond in different ways. The total time-lags are the result of a combination of both the source and disc responses. Their model predictions appear to be in agreement (qualitatively) with the peaky residuals in our Fig.~\ref{fig:frequency} (see their Fig.~4). We plan in the future to implement an option to {\tt KYNREFREV} to account for a spatially extended azimuthally symmetric corona in a few simplified configurations. 

\section*{Acknowledgments}

MCG, MD and MB acknowledge support provided by the European Seventh Frame-work 
Programme (FP7/2007-2013) under grant agreement n$^{\circ}$ 312789 and GA CR grant 18-00533S.

\bsp

\label{lastpage}

\vspace{-0.5cm}
\begin{thebibliography}{99}

\bibitem[Alston et al.(2013)]{alston13}
Alston, W.~N., Vaughan, S. \& Uttley, P., 2013, MNRAS, 435, 1511
\bibitem[Ar{\'e}valo et al.(2006)]{arevalo06} Ar{\'e}valo, P., Papadakis, I.~E., Uttley, P. et al.(2006), MNRAS, 372, 401
\bibitem[Arnaud et al.(1996)]{arnaud96} Arnaud, K.~A. (1996), Astronomical Society of the Pacific Conference Series, 101, 17
\bibitem[Bentz et al.(2016)]{bentz16} Bentz, M.~C., Cackett, E.~M., Crenshaw, D.~M. et al., (2016), ApJ, 830, 136
\bibitem[Boller et al.(2002)]{boller02}
Boller, T., Fabian, A.~C., Sunyaev, R. et al., 2002, MNRAS, 329L, 1
\bibitem[Brenneman \& Reynolds (2006)]{brenneman06} Brenneman, L.~W. \& Reynolds, C.~S. (2006) ApJ, 652, 1028
\bibitem[Cackett et al.(2014)]{cackett14} Cackett, E.~M., Zoghbi, A., Reynolds, C. et al., 2014, MNRAS, 438, 2980
\bibitem[Chainakun et al.(2016)]{chainakun16}
Chainakun, P., Young, A.~J. \& Kara, E., 2016, MNRAS, 460, 3076
\bibitem[Chainakun et al.(2017)]{chainakun17}
Chainakun, P. \& Young, A.~J., 2017, MNRAS, 465, 3965
\bibitem[Dauser et al.(2012)]{dauser12}
Dauser, T., Svoboda, J., Schartel, N. et al., 2012 MNRAS, 422, 1914
\bibitem[De Marco et al.(2009)]{demarco09}
De Marco, B., Iwasawa, K., Cappi, M., et al., 2009, A\&A, 507, 159
\bibitem[De Marco et al.(2013)]{demarco13} De Marco, B., Ponti, G., Cappi, M. et al. (2013), MNRAS, 431, 2441
\bibitem[Done et al.(2016)]{done16} 
Done, C. \& Jin, C., 2016, MNRAS, 460, 1716
\bibitem[Dov{\v c}iak et al.(2014)]{dovciak14}
Dovciak, M., De Marco, B., Kara, E. et al., 2014, The X-ray Universe 2014, 244
\bibitem[Dov{\v c}iak et al.(2018, in prep.)]{dovciak18}
Dovciak et al., in prep., 2018
\bibitem[Emmanoulopoulos et al.(2011)]{emmanoulopoulos11} Emmanoulopoulos, D., McHardy, I.~M., Papadakis, I.~E. (2011) MNRAS, 416, 94
\bibitem[Emmanoulopoulos et al.(2014)]{emmanoulopoulos14} Emmanoulopoulos, D., Papadakis, I.~E., Dov{\v c}iak, M. \& McHardy, I.~M., 2014, MNRAS, 439, 3931
\bibitem[Epitropakis et al.(2016a)]{epitropakis16a} Epitropakis, A., Papadakis, I.~E., Dov{\v c}iak, M. et al., 2016a, A\&A, 594, 71
\bibitem[Epitropakis \& Papadakis(2016b)]{epitropakis16b} Epitropakis, A. \& Papadakis, I.~E., 2016b, A\&A, 591A, 113
\bibitem[Epitropakis \& Papadakis(2017)]{epitropakis17} Epitropakis, A. \& Papadakis, I.~E., 2017, MNRAS, 468, 3568
\bibitem[Fabian et al.(1999)]{fabian99} Fabian, A.~C., Celotti, A., Pooley, G. et al. (1999) MNRAS, 308, 6
\bibitem[Fabian et al.(2003)]{fabian03}
Fabian, A.~C. \& Vaughan, S., 2003, MNRAS, 340L, 28
\bibitem[Fabian et al.(2004)]{fabian04}
Fabian, A.~C., Miniutti, G., Gallo, L. et al., 2004, MNRAS, 353, 1071
\bibitem[Fabian et al.(2009)]{fabian09} Fabian, A.~C., Zoghbi, A., Ross, R.~R. et al., 2009, Nature, 459, 540
\bibitem[Fabian et al.(2012)]{fabian12} Fabian, A.~C., Zoghbi, A., Wilkins, D. et al., 2012, MNRAS, 419, 116 
\bibitem[Gabriel et al.(2004)]{gabriel04} 
Gabriel C., et al., 2004, in Ochsenbein F., Allen M. G., Egret D., eds., Astronomical Society of the Pacific Conference Series Vol. 314, Astronomical Data Analysis Software and Systems (ADASS) XIII. p. 759
\bibitem[Giustini et al.(2015)]{giustini15}
Giustini, M., Turner, T.~J., Reeves, J.~N. et al., 2015, A\&A, 577, 8
\bibitem[Guainazzi et al.(1999)]{guainazzi99}
Guainazzi, M., Matt, G., Molendi, S. et al., 1999, A\&A, 341L, 27
\bibitem[Kara et al.(2013a)]{kara13a}
Kara, E., Fabian, A.~C., Cackett, E.~M., et al., 2013, MNRAS, 428, 2795
\bibitem[Kara et al.(2013b)]{kara13b}
Kara, E., Fabian, A.~C., Cackett, E.~M. et al., 2013, MNRAS, 434, 1129
\bibitem[Kara et al.(2014)]{kara14}
Kara, E., Fabian, A.~C., Marinucci, A. et al., 2014, MNRAS, 445, 56
\bibitem[Kara et al.(2016)]{kara16}
Kara, E., Alston, W.~N., Fabian, A.~C., et al., 2016, MNRAS, 462, 511
\bibitem[Kara et al.(2017)]{kara17}
Kara, E., Garc{\'{\i}}a, J.~A., Lohfink, A., et al., 2017, MNRAS, 468, 3489
\bibitem[Kaspi et al.(2000)]{kaspi00}
Kaspi, S., Smith, P.~S., Netzer, H. et al., 2000, ApJ, 533, 631
\bibitem[Laha et al.(2014)]{laha14}
Laha, S., Guainazzi, M., Dewangan, G.~C. et al., 2014, MNRAS, 441, 2613
\bibitem[Lee et al.(2001)]{lee01}
Lee, J.~C., Ogle, P.~M., Canizares, C.~R. et al., 2001, ApJ, 554L, 13
\bibitem[Marinucci et al.(2014)]{marinucci14}
Marinucci, A., Matt, G., Miniutti, G. et al., 2014, ApJ, 787, 83
\bibitem[Martocchia \& Matt (1996)]{martocchia96} Martocchia, A. \& Matt, G. (1996), MNRAS, 282L, 53
\bibitem[Mastroserio et al.(2018)]{mastroserio18} Mastroserio, G., Ingram, A. \& van der Klis, M., 2018, MNRAS, 475, 4027
\bibitem[McHardy et al.(2004)]{mchardy04} McHardy, I.~M., Papadakis, I.~E., Uttley, P. et al.(2004), MNRAS, 348, 783
\bibitem[McHardy et al.(2007)]{mchardy07} McHardy, I.~M., Ar{\'e}valo, P., Uttley, P., et~al.(2007), MNRAS, 382, 985
\bibitem[Miller et al.(2007)]{miller07} Miller, L., Turner, T.~J., Reeves, J.~N. et al., 2007, A\&A, 463, 131
\bibitem[Miniutti et al.(2007)]{miniutti07} 
Miniutti, G., Fabian, A.~C., Anabuki, N. et al., 2007, PASJ, 59S, 315
\bibitem[Mizumoto et al.(2014)]{mizumoto14} 
Mizumoto, M., Ebisawa, K. \& Sameshima, H., 2014, PASJ, 66, 122
\bibitem[Miyamoto \& Kitamoto (1989)]{miyamoto89} Miyamoto, S. \& Kitamoto, S., (1989) Nature, 342, 773
\bibitem[Nayakshin et al.(2000)]{nayakshin00} Nayakshin, S., Kazanas, D. \& Kallman, T.~R., 2000, ApJ, 537, 833 
\bibitem[Nowak \& Vaughan (1996)]{nowak96} Nowak, S. \& Vaughan, S., (1996) MNRAS, 280, 227
\bibitem[Nowak et al.(1999)]{nowak99} Nowak, M.~A., Wilms, J. \& Dove, J.~B., (1999) MNRAS, 517, 355
\bibitem[Otani et al.(1996)]{otani96}
Otani, C., Kii, T., Reynolds, C.~S., et al., 1996, PASJ, 48, 211
\bibitem[Pan et al.(2016)]{pan16}
Pan, H.-W., Yuan, W., Yao, S. et al., 2016, ApJ, 819L, 19
\bibitem[Papadakis et al.(2001)]{papadakis01} Papadakis, I.~E., Nandra, K. \& Kazanas, D. (2001), ApJ, 554, 133
\bibitem[Papadakis et al.(2007)]{papadakis07}
Papadakis, I.~E., Brinkmann, W., Page, M.~J. et al., 2007, A\&A, 461, 931
\bibitem[Ross \& Fabian (2005)]{rossfabian05} Ross, R.~R. \& Fabian, A.~C., 2005, MNRAS, 358, 211
\bibitem[Silva et al.(2016)]{silva16} Silva, C.~V., Uttley, P. \& Costantini, E., 2016, A\&A, 596, 79
\bibitem[Svoboda et al.(2012)]{svoboda12} Svoboda, J., Dov{\v c}iak, M., Goosmann, R.~W., et~al., 2012, A\&A, 545, 106
\bibitem[Tanaka et al.(1995)]{tanaka95}
Tanaka, Y., Nandra, K., Fabian, A.~C. et al., 1995, Natur, 375, 659
\bibitem[Tombesi et al.(2010)]{tombesi10}
Tombesi, F., Cappi, M., Reeves, J.~N. et al., 2010, A\&A, 521, 57
\bibitem[Turner et al.(2003)]{turner03} Turner, A.~K., Fabian, A.~C., Vaughan, S. et al., 2003, MNRAS, 346, 833
\bibitem[Turner et al.(2004)]{turner04} Turner, A.~K., Fabian, A.~C., Lee, J.~C. \& Vaughan, S., 2004, MNRAS, 353, 319
\bibitem[Turner et al.(2007)]{turner07} Turner, T.~J., Miller, L., Reeves, J.~N., Kraemer, S.~B., 2007, A\&A, 475, 121
\bibitem[Vaughan et al.(2004)]{vaughan04}
Vaughan, S. \& Fabian, A.~C., 2004, MNRAS, 348, 1415 
\bibitem[Vestergaard \& Peterson (2006)]{vestergaard06} Vestergaard, M. \& Peterson, B.~M., 2006, ApJ, 641, 689
\bibitem[Wilms et al.(2001)]{wilms01}
Wilms, J., Reynolds, C.~S., Begelman, M.~C. et al., 2001, MNRAS, 328L, 27
\bibitem[Zoghbi et al.(2010)]{zoghbi10}
Zoghbi, A., Fabian, A.~C., Uttley, P. et al., 2010, MNRAS, 401, 2419
\bibitem[Zhou \& Wang (2005)]{zhou05} Zhou, X.-L., Wang, J.-M., 2005, ApJ, 618L, 83


\end{thebibliography}
\end{document}